\documentclass[aps,floatfix,showkeys,superscriptaddress,notitlepage]{
revtex4}
\usepackage{amsmath,amssymb,amsfonts,graphics,graphicx,dcolumn,bm,enumerate}
\usepackage{comment,natbib,appendix}
\usepackage{multirow,color}

\newcommand{\sinp}
{\affiliation{Condensed Matter Physics Division, 
Saha Institute of Nuclear Physics, 1/AF Bidhannagar, Kolkata 700064, India.}}
\newcommand{\iima}
{\affiliation{Economics area, Indian Institute of Management, Vastrapur, 
Ahmedabad 380015, India}}
\newcommand{\aalto}
{\affiliation{Department of Computer Science, Aalto University School of 
Science,
P.O. Box 15400, FI-00076 AALTO, Finland}}
\newcommand{\jnu}
{\affiliation{School of Computational and Integrative Sciences, Jawaharlal Nehru 
University, New Delhi 110067, India}}
\newcommand{\csss}
{\affiliation{CTRPFP, Centre for Studies in Social Sciences, Calcutta, 
R–1 Baishnabghata Patuli Township, Kolkata 700094, India}}

\begin{document}

\title{Invariant features of spatial inequality in consumption: the case of 
India}

\author{Arnab Chatterjee}%
\email[Email: ]{arnabchat@gmail.com}
\sinp
\author{Anindya S. Chakrabarti}
\email[Email: ]{anindyac@iimahd.ernet.in}
\iima 
\author{Asim Ghosh}
\email[Email: ]{asimghosh066@gmail.com}
\aalto
\author{Anirban Chakraborti}
\email[Email: ]{anirban@jnu.ac.in}
\jnu
\author{Tushar K. Nandi}
\email[Email: ]{nandi.tushar@gmail.com}
\csss

\begin{abstract}
  
We study the distributional features and inequality of consumption expenditure 
across India, for different states, castes, religion and urban-rural divide. 
We find that even though the aggregate measures of inequality are fairly 
diversified across states, the consumption distributions show near identical 
statistics, once properly normalized. This feature is seen to be robust with 
respect to variations in sociological and economic factors.
We also show that state-wise inequality seems to be positively correlated with 
growth which is in accord with 
the traditional idea of Kuznets' curve.
We present a brief model to account for the invariance found empirically and 
show that better but riskier technology draws can create a positive correlation 
between inequality and growth.

\end{abstract}

\maketitle

\section{Introduction}
\label{sec:intro}
Socio-economic 
inequality~\cite{arrow2000meritocracy,
stiglitz2012price,chatterjee2015socio-economic,chatterjee2015socialinequality}
has been a subject of high interest, and is drawing much attention recently. 
Economic inequality can be measured in terms of three major variables: wealth, 
income and consumption. For decades, if not centuries, economists (and very 
recently, physicists)
have worked on the statistical descriptions, 
manifestations and precise mechanisms giving rise to such inequalities  
\cite{piketty2014capital,chakrabarti2013econophysics,Cohen;14}. Income 
inequality is the most cited measure followed by
wealth inequality which in turn is followed by consumption inequality. Such ordering arises primarily because of availability of data (or lack of it). From tax documents collected by the government, it is somewhat easier
to figure out the income distribution rather than, say, keeping track of individual consumer expenditures. Thus most of the literature has skipped the consumption
inequality part \cite{Blundell;08}. However, from the households' perspective that appears to be the most important factor. Both income and wealth are vehicles through which the ultimate goal of consumptions are
satisfied. Thus even though it is more difficult to reliably measure and 
quantify, consumption inequality is a much better proxy for social/economic 
welfare than the others.
In developing economies with limited coverage of direct tax system, official 
data on income represent a very small proportion on population. Rather 
consumption data in such context is more reliable and offer greater coverage.

Interestingly, there are a number of very well known and established regularities in income and wealth distributions.
Pareto~\cite{Pareto-book} made extensive studies by the end of the 19th 
century, and found that wealth distribution in Europe follows a power law 
for the richest, later came to be known as the \textit{Pareto law}.
Subsequent studies revealed that the distributions of income 
and wealth possess a number of fairly robust features:
the bulk of both the income and wealth distributions seem to reasonably fit both 
the log-normal and the Gamma distributions (see, e.g., 
\cite{chakrabarti2013econophysics}). Economists prefer the 
log-normal distribution~\cite{Montroll:1982,gini1921measurement}, while 
statisticians~\cite{Hogg-2007} and 
physicists~\cite{druagulescu2001exponential,Chatterjee:EWD,Chatterjee-2007,
Banerjee:2010,Yakovenko:RMP} 
emphasize on the Gamma distribution for the probability density or Gibbs/ 
exponential distribution for the corresponding cumulative distribution. 
However, the high end of the distribution (known as the `tail') fits well 
to a power law as observed by Pareto, the exponent known as the
Pareto exponent, usually ranging between 1 and 
3~(see e.g.~\cite{chakrabarti2013econophysics}; 
Ref.~\cite{Wileybook2} contains a historical account of Pareto's data as well 
as some recent sources). 
For India, the wealthiest have been found to have their assets distributed along
a power law tail~\cite{sinha2006evidence,jayadev2008power}.
There has even been a study on per capita energy 
consumption~\cite{lawrence2013global} which 
shows that the global inequality in that respect is decreasing.

Although income and wealth distribution data are used to quantify 
economic inequality for individuals or family/households, distribution of 
consumer expenditure should also reflect certain aspects of disparity in 
society. However, such studies are comparatively  rare compared to 
the much commonly reported studies of income and wealth, mostly due to the 
difficulty in acquiring data.
A recent study reported the expenditure of individuals in a single shopping
bill, using data from bills corresponding to a chain of a particular 
convenience store in Japan~\cite{mizuno2008pareto}. The probability density of 
expenditure was found to have a power law tail with exponent $2$, and the Gini
index was found to be around $0.70$.
Another study~\cite{battistin2009consumption}
found that the household expenditure distribution quite close to log-normal for 
US and UK.
A rigorous study on the household consumer expenditure in 
Italy~\cite{fagiolo2010distributional}
reported that the distribution function is not a lognormal but ``invariably 
characterized by asymmetric exponential power densities''.
A very recent work~\cite{ghosh2011consumer} reported that the tail of the 
consumer expenditure in India follows a power law distribution along 
with a lognormal bulk, in the same way as income distribution does.

Statistical physics presents an idea of `universality', where a system of many 
interacting dynamical units collectively exhibit a behavior, which simply 
depends on only a few basic (dynamical) features  of the individual constituent 
units, and the embedding dimension of the system, but is independent of all 
further details.
Socio-economic data exhibit enough empirical evidences in support of 
universality, which prompt a community of researchers to propose simple,
minimalistic models to understand them, similar to those commonly used in 
statistical physics. Typical examples are 
elections~\cite{fortunato2007scaling,chatterjee2013universality}, 
population growth~\cite{rozenfeld2008laws} and 
economy~\cite{stanley1996scaling},
income and wealth distributions~\cite{chakrabarti2013econophysics},
languages~\cite{petersen2012statistical}, etc.
(see Refs.~\cite{Castellano:2009,Sen:2013} for reviews).

In this paper, we present a statistical analysis of consumption profiles of 
households across all states of India. 
We show that the bulk of the data can be described well by a lognormal distribution. What is even more intriguing is that the cross-sectional distributions
of consumption expenditure show near identical distributional features under suitable normalization. Thus effectively different states are different in terms of scale only
and not in terms of other factors which can potentially create dispersion in 
consumption across population, supporting a `universality' hypothesis. 
Furthermore, we see that increasing per head consumption (which is a good proxy 
of per head income) leads to higher inequality.
We present a vary basic model to illustrate the mechanism which shows that why increasing the size of the pie can potentially lead to more inequality. This also explains the
invariance in inequality with respect to proper scaling.

\section{Data description}
\label{sec:datadesc}
We use the data for Household Consumer Expenditure 66th Round from the 
National Sample Survey Office (NSSO)~\cite{NSSO}. The data contains information 
about expenditure incurred by households on consumption goods and services 
during the reference period. In general, these sample surveys are conducted using households as unit of the economy. This leads to a problem because of heterogeneity in
household size and this dispersion in size is often quite big. To take that into account, both 
household and members of the household were 
counted and proper normalization methods are used.

Data is available for all sampled households in the different states and 
Union territories (UT), across several parameters like castes, religions and 
rural-urban divide.
We use the NSSO data set for the year 2009-2010 on consumption expenditure.
There are 100957 households in this data set.
To study the inequality structure, we use two kinds of data which provides two perspectives.
The first one is the monthly per capita consumer expenditure 
(MPCE) which is simply the total consumption expenditure of a household per 
household member. The second one being the monthly per capita equivalent 
consumption expenditure (MPECE).
In the former (MPCE), all the members of the household are assumed to have the 
same weight, while in the latter (MPECE), household members are given different 
weights according to their age, i.e., adults get a higher weight than a 
child~\cite{hagenaars1996poverty}.

\section{Measuring inequality}
\label{sec:measureineq}
To quantify the degree of inequality, we use two measures. The first one is the 
standard Gini coefficient. 
This happens to be the most popular methods of measuring inequality. One  
considers the Lorenz curve, which  represents 
the cumulative proportion $X$ of ordered (from lowest to highest) individuals 
(entries) in terms of the cumulative proportion of their sizes $Y$.
$X$ can represent income or wealth of individuals, and in our case it is 
household consumer expenditure.
But it can as well be citation, votes, city population etc. of articles, 
candidates, cities respectively (see e.g. Ref. \cite{Satya;09}).
The Gini index ($g$) is defined as the ratio between the area enclosed between 
the Lorenz curve and the equality line, to that below the equality line.
It is the most common measure to quantify socio-economic inequality.
Ghosh et al.~\cite{ghosh2014inequality}
recently introduced the `$k$ index', which is 
defined as the fraction $k$ such that  $(1-k)$ fraction 
of people  possess $k$ fraction of income or 
expenditure~\cite{inoue2015measuring}.

 \section{Results}
 \label{sec:results}
We compute distribution of the variable $x$ which is the consumption 
expenditure for the two sets (MPCE \& MPECE). 
Ref. \cite{Ravallion;98} shows that the Indian states have progressed very differently over time in terms of alleviating poverty at the grass-root level. 
The general conclusion drawn from this work is that basically those who started with better infra-structure and bigger human-capital stock, did better in pulling people out of
misery. What is of importance is that a major determinant in the whole process was their dependence on external macroeconomic factors rather than idiosyncratic micro ones. Thus in short-run relative prosperity depends
a lot on the inflationary outcomes whereas technological shocks have pronounced and adverse effects. 

Ref. \cite{Mishra;92} was one of the very first attempts to quantify the degree of inequality at the state level in case of India and calculate its contribution to the national inequality.
They found that urban-rural inequality explains a substantial part of the state-level inequality which in turn, explains national inequality. Thus there was large amount of within state inequality
around 1980s (their data coverage was 1977-78 and 1983). We show that while there is substantial inter-state (between sates) inequality in the present data set (2009-10), intra-state (within states) measures of inequality have surprisingly similar statistical features.

\subsection{Invariance in consumption expenditure}
\label{Subsec:inv_cons}
\begin{figure}
 \includegraphics[width=\linewidth]{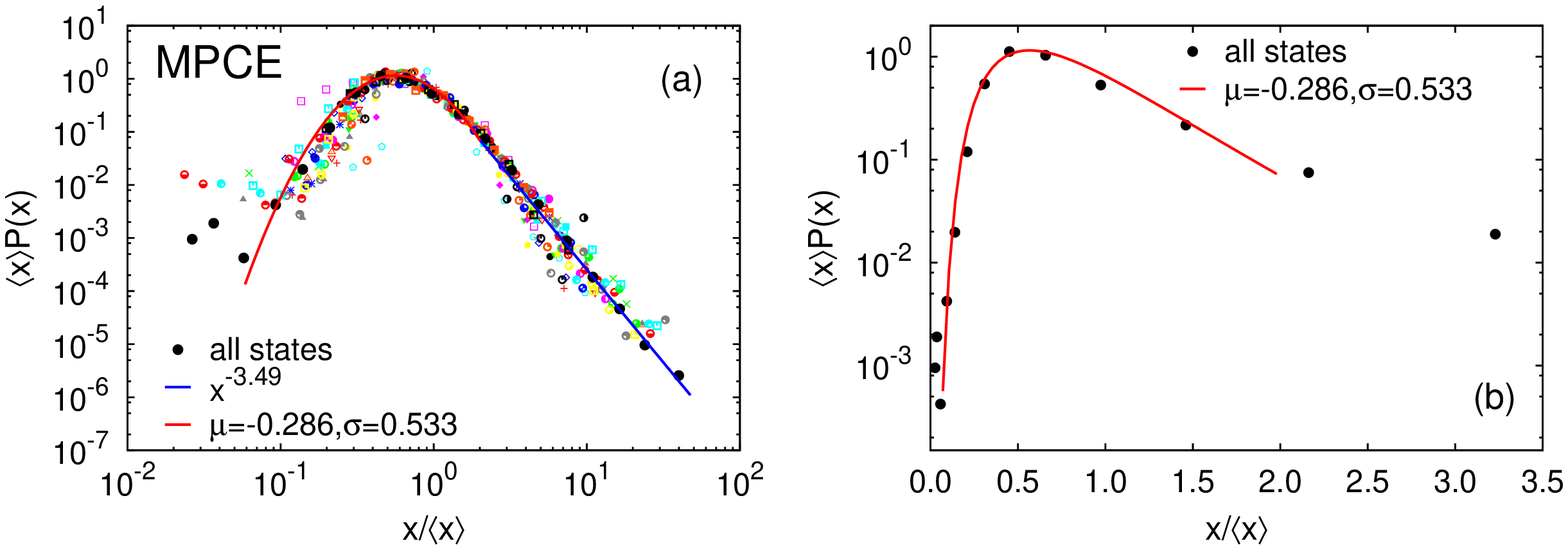}
  \includegraphics[width=\linewidth]{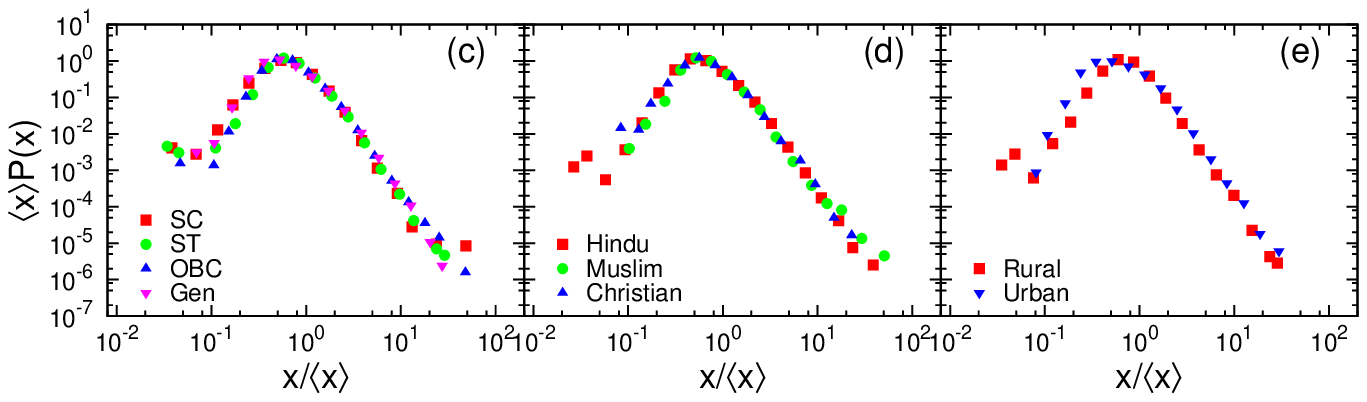}
 \caption{Data for monthly per capita consumer expenditure 
(MPCE),
 showing the probability density $P(x)$ of per capita household 
expenditure $x$ for 2009-2010. 
The data is rescaled by averages $\langle x \rangle$. 
  (a) The black solid circle represents the data 
for all states, while other symbols indicate each of the 35 individual states 
and UTs.
The lognormal fit to the bulk of the distribution has parameters 
$\mu=-0.286, \sigma=0.533$ while the power law tail has decay exponent $3.49$;
(b) The same data for all states plotted in log-linear scale, 
with the lognormal
curve for comparison.
(c) for 4 different caste tags - SC, ST, OBC and General;
(d) for 4 different religion tags - Hindu, Muslim, Christian and Others;
(e) for rural and urban population.
}
 \label{fig:states}
\end{figure}

\begin{figure}
 \includegraphics[width=\linewidth]{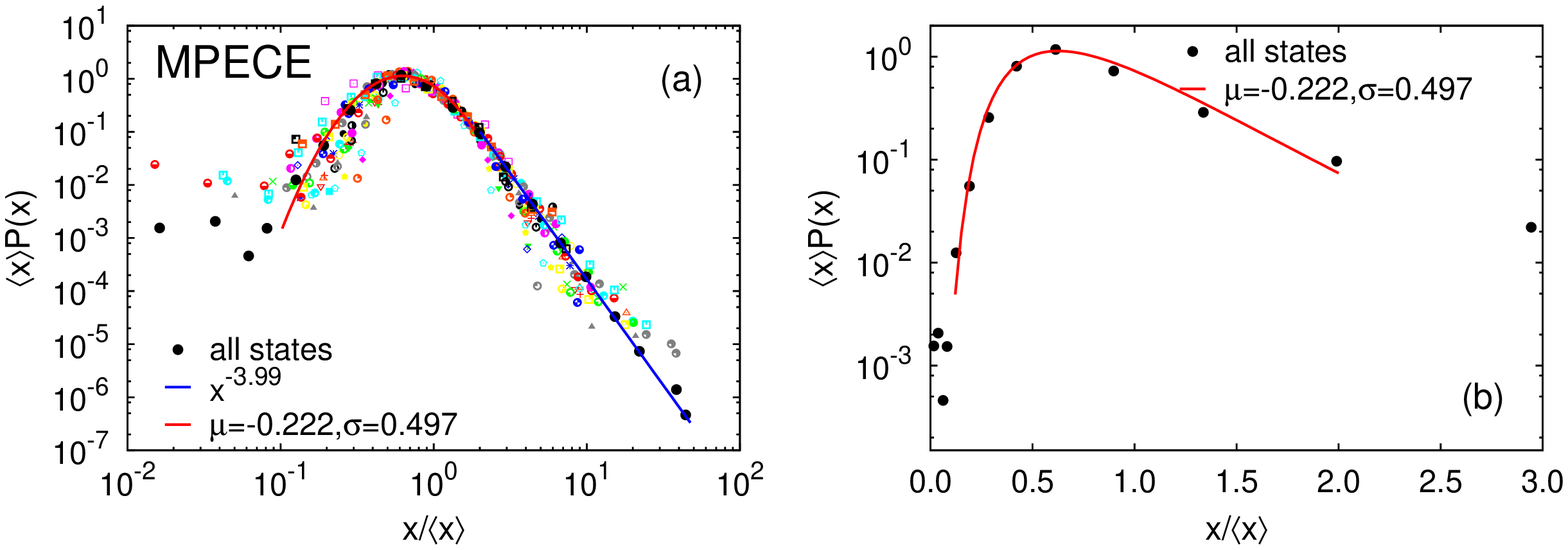}
 \includegraphics[width=\linewidth]{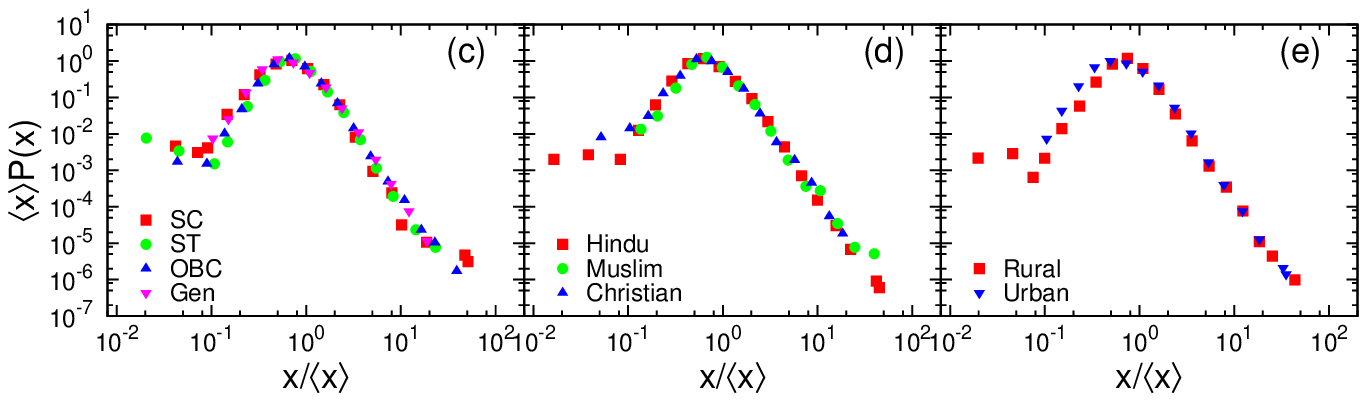}
 \caption{Data for monthly per capita equivalent consumer 
expenditure (MPECE), showing the probability density $P(x)$ of per capita 
household 
expenditure $x$ for 2009-2010. 
The data is rescaled by averages  $\langle x \rangle$. 
  (a) The black solid circle represents the data 
for all states,  while other symbols indicate each of the 35 individual states 
and UTs.
The lognormal fit to the bulk of the distribution has parameters 
$\mu=-0.222, \sigma=0.497$ while the power law tail has decay exponent $3.99$;
(b) The same data for all states plotted in log-linear scale, 
with the lognormal
curve for comparison.
(c) for 4 different castes -- SC, ST, OBC and General;
(d) for 4 different religion tags -- Hindu, Muslim, Christian and Others;
(e) for rural and urban population.
}
 \label{fig:Estates}
\end{figure}

We first compute the probability density $P(x)$ (left panels) and then rescale 
with the mean ($\langle x \rangle = \mu$) for a possible data collapse. 
In Figs. \ref{fig:states}(a) and \ref{fig:Estates}(a), we present the 
data for all states for MPCE \& MPECE respectively.
While we acknowledge the fact that the sample size is small for 
certain states 
(like Goa, Andaman), we show that the pattern holds true for all states. Also 
other sociological or geographic factors do not affect it, for which we have 
significantly large number of data points.
We also plotted the full data (black filled circles) and estimated best fits.
The bulk of the distribution fits well to a lognormal 
$P(x) = \frac{1}{x \sigma \sqrt{2 \pi}} \exp \left[ -\frac{(\ln x- \mu)^2}{2 
\sigma^2} \right]$. 
For MPCE, the parameters are $\mu=-0.286, \sigma=0.533$ while for
MPECE  parameters are $\mu=-0.222, \sigma=0.497$.
However, for the lowest values, the lognormal fit does not 
hold.
The largest values of consumption expenditure fit well to power law
$ax^{-b}$. 
For MPCE, the decay exponent is $b=3.49$ and for MPECE, $b = 3.99$.
In Figs. \ref{fig:states}(b) and \ref{fig:Estates}(b), we plot 
the binned full data and the lognormal fit for comparison in log-linear scale.
Subsequently, we also show the data filtered according to 3 available 
parameters: caste (Figs.~\ref{fig:states}(c) and \ref{fig:Estates}(c)), 
religion (Figs.~\ref{fig:states}(d) and \ref{fig:Estates}(d)) and rural-urban 
divide (Figs.~\ref{fig:states}(e) and \ref{fig:Estates}(e)).
Again the data show excellent scaling collapse, indicating that the basic 
functional form of the probability distribution $P(x)$  is invariant
with respect to different states (spatial invariance), caste, religion or 
rural-urban divide. 
We also checked that lognormals are the best fits for the 
rescaled plots (as in panel (b) of Figs.~\ref{fig:states} and 
\ref{fig:Estates}; not shown here).

\subsection{Rescaling the data}
\begin{figure}
\begin{center}
 \includegraphics[width=\linewidth]{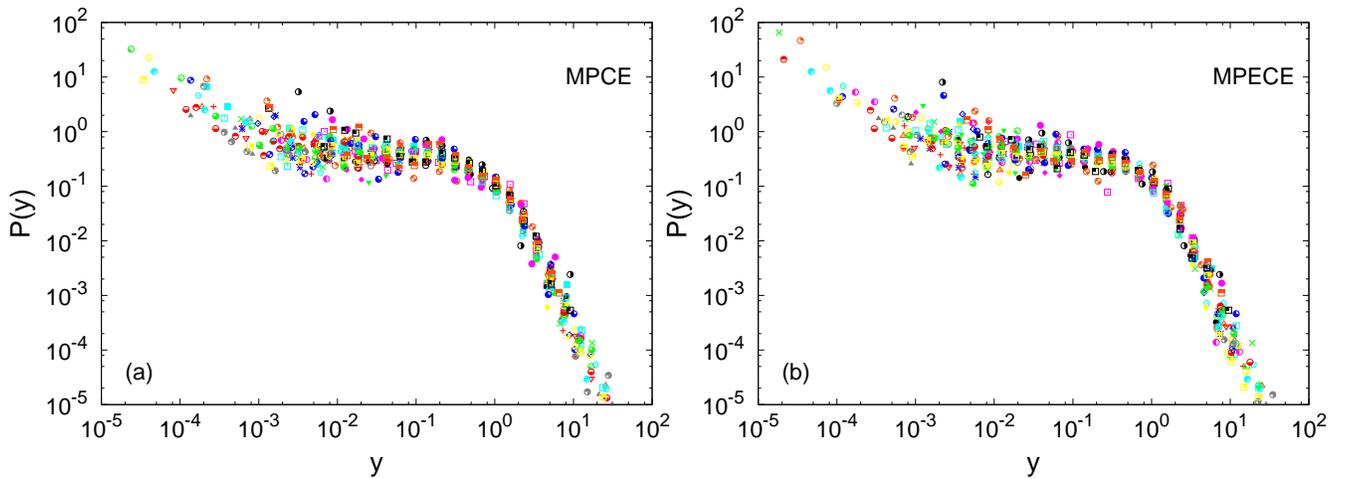}
\caption{The normalized probability density $P(y)$ vs. 
$y=\frac{x-\mu}{\sigma}$, where $\mu$ is the sample mean of $x$ and $\sigma$ is 
the sample standard deviation.
Data for per capita household 
expenditure $x$ for 2009-2010.
Each type of symbol correspond to one of 35 different states or UTs.
Data is for (a)  MPCE and (b) MPECE.
 }
 \label{fig:normal}
\end{center}
 \end{figure}

In section \ref{Subsec:inv_cons}, the scaling was done with respect to the 
average income $\mu=\langle x \rangle$. Here another type of rescaling is 
presented. 
We compute the distribution for the consumption expenditure 
data  normalized with respect to the mean and standard deviation; 
$y=\frac{x-\mu}{\sigma}$, where $\mu$ is the sample mean of $x$ and $\sigma$ is the sample standard 
deviation. See Figs. \ref{fig:normal} for the data collapse in MPCE and MPECE. 

Table~\ref{tab:per_cap_exp_gini_1} compile 
data on per capita expenditure across all states and their corresponding Gini 
coefficients and $k$-index (see App. \ref{subsec:appendix} for a complete description).
 
\subsection{Inequality along other social dimensions}
 A basic proposition of this paper is that the distributional features of 
consumption expenditure are very similar subject to
normalization. So far we have considered spatial dimension of consumption inequality.
There are interesting features of inequality in terms of other social factors. Labor economists have 
shown that wage structure varies considerably across spectrum of sex, age and racial backgrounds \cite{Hurst;13}. 
In the present context, a robust feature comes out when we consider similar factors. Surprisingly,
inequality in consumption profile across various religious groups and ethnic groups are near identical
subject to a scaling factor.

It is important to recognize that income is not the only determinant of well-being even though it has the basic virtue of
being easily quantifiable and hence, comparable. The literature has focused on poverty from a multi-dimensional perspective
incorporating various other factors \cite{Satya;forthcoming}. Its theoretical support comes form the capability approach presented by \cite{Sen-capability;99}
which argues that poverty is manifestation of failure of a person to exercise his/her capabilities to the extent possible.
The reasons why such failures exist come in various forms. Religious and ethnic backgrounds constitute two extremely important factors
in terms of social and economic barriers. Interestingly we find that the basic statistical features do not change much after normalizing the data. This, in principle, reflects that
to understand consumption inequality we need a model that generates dispersion in consumption and consumption profiles need to be multiplicative in nature so that subject to scaling, it generates
identical patterns.

Fig. \ref{fig:states}(c) shows the normalized probability density functions for 
data compiled conditional on castes. Next, we study the dispersion in 
consumption expenditure across religions.
Fig. \ref{fig:states}(d) clearly shows that the distributional features are 
very similar across religions. Finally, we have studied dispersion in 
expenditure across urban versus rural economy. 
Fig. \ref{fig:states}(e) shows that under normalization, similar features 
prevail. This finding needs some elaboration. As noted above, there is a large 
literature in labor economics exploring the
gap in consumption across various social and ethnic groups and of course, the urban-rural consumption gap has been recognized for long. What these findings suggest is that we do not need different models to explore the dispersion under different conditions. In other words, while it is the case that the economic pie is bigger for some group of people (for example, urban) than their counterpart
(say, rural), that indicates absolute inequality. Relative inequalities are of 
similar nature in both cases. We exploit this property below when we try to come 
up with a coherent version of the broad picture.

Similar findings persist when we do the same exercise for the MPECE dataset. See 
Fig. \ref{fig:Estates}.

\subsection{Comparison between the two data sets}
\begin{figure}[t]
 \includegraphics[width=\linewidth]{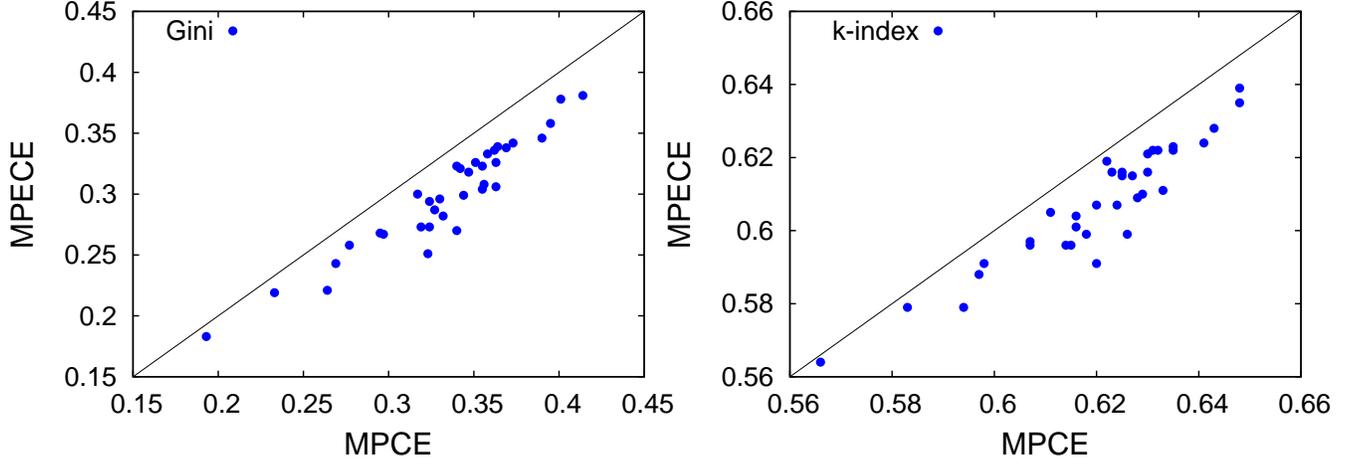}
\caption{Comparisons between Gini and $k$ indices from the two data sets across the states. MPCE consistently shows more inequality than MPECE as all points without exception lies below the $45^{\circ}$ line.}
 \label{fig:gvsg}
 \end{figure}

Fig.~\ref{fig:gvsg} compares the values of Gini and the $k$-index as computed 
from the two data sets.
Gini index ranges approximately between $0.19 - 0.41$ for MPCE, 
$0.18 - 0.38$ for MPECE while for 
k-index, it is $0.56-0.65$ for MPCE and $0.56 - 0.64$ for MPECE.

Clearly, MPCE consistently shows more inequality than MPECE in terms of both measures.
It is worthwhile to reiterate the definitions here. 
MPCE is just monthly expenditure on consumption per head of a family. It is 
calculated as the total consumption expenditure of a household divided by the 
number of household members. 
It is important to note that each household member gets equal weight in this 
formula.
On the other hand, MPECE takes into account intra-family heterogeneity. It is 
monthly per capita equivalent consumption expenditure which is calculated 
assigning different weights to household members depending on their age. In 
particular, an adult member gets a higher weight than a child.
The main purpose of using equivalence scale is to account for the consumption 
of goods for common use, like fuel, accommodation etc. Consumption of these 
goods does to necessarily grow in proportion to household size. Rather, age 
structure plays crucial role in the consumption.

To see why MPCE is more unequal than MPECE, consider an island $k$ with $N$ families with size sequence $\{s_1,s_2,\ldots s_N\}$ i.e. the $i$-th family has size $s_i$. 
To maintain clarity, below the quantities do not carry the island index with the understanding that we are talking only about the 
$k$-th island. When we make comparisons between islands, we will start indexing the islands by $k$.
Let us denote the total expenditure for each family by $\{E_1, E_2,\ldots ,E_N\}$ where the total expenditure is the sum of all individual expenditures on the family members,
\begin{equation}
E_n=\sum^{s_n}_i e_{in}.
\end{equation}
Without any loss of generality, let us also assume that the expenditures are ranked so that 
\begin{equation}
E_1 \ge E_2 \ge \ldots \ge E_N.
\label{Eqn:reorderE}
\end{equation} 
By definition, we have the MPCE as of the $n-$th family as
\begin{equation}
MPCE_n=\sum^{s_n}_i \frac{1}{s_n}.e_{in} 
\label{Eqn:mpce}
\end{equation}
where all family members get the same weight $1/s_n$
and similarly, we can write MPECE as a weighted average of individual 
expenditures with weights $\{w_i\}$ for the $i$-th member,
\begin{eqnarray}
MPECE_n &=& \sum^{s_n}_i w_i.e_{in},  \nonumber \\
       &=& \sum^{s_n}_i \frac{1}{s_n}.e_{in} +\sum^{s_n}_i (w_i-\frac{1}{s_n}).e_{in}, \nonumber \\
			 &=& MPCE_n + \epsilon_n
\label{Eqn:mpce=mpece}
\end{eqnarray}
where in the last line we employed Eq.~\ref{Eqn:mpce} and the last term 
$\epsilon_n$ combines the dispersion of expenditures from the average.
It is noteworthy here that in general,
\begin{equation}
\epsilon_n \ge 0.
\end{equation}
The reason is that in the definition of MPECE, the adults get a higher weight ($w_{adult}>w_{child}$) and typically the individual expenditure on them would also be high i.e. $e_{adult} > e_{child}$. Thus there is an upward bias in $\epsilon_n$. This has an immediate corollary which is
\begin{eqnarray}
\langle MPECE_n\rangle &=& \langle MPCE_n \rangle + \langle \epsilon_n \rangle \nonumber \\
        & \ge & \langle MPCE_n \rangle 
\label{Eqn:mpce_mpece}
\end{eqnarray}
which means that for any island the average MPECE would be larger than average 
MPCE. This can be verified easily from Table~\ref{tab:per_cap_exp_gini_1}.
Note that Gini coefficient of an island with per capita expenditure 
profile $\{\tilde{e}_{n}\}$ (where $\tilde{e}$ is either MPCE and MPECE) across families $n\in N$ can be written as
\begin{eqnarray}
G &=& 1+\frac{1}{N}- \left( \frac{2}{N^2 \langle \tilde{e}_{n} \rangle} \right) [\sum_n^N n.\tilde{e}_{n}], \nonumber \\
    &=& 1+\frac{1}{N}- \left(\frac{2}{N^2}\right) .X ~~~\mbox{say.}
\label{Eqn:gini_expand}
\end{eqnarray}
Now note that, we have (ignoring the index for the island)
\begin{eqnarray}
X^{MPECE} &=&  \frac{\sum_n^{N} n.\tilde{e}_{n}}{\langle \tilde{e}_{n} \rangle}, \nonumber \\
    &=& \frac{\sum_n^{N} n. MPECE_{n}}{\frac{1}{N}\sum_n^{N} MPECE_n } , \nonumber \\
		&=& N.\left(\frac{\sum_n^{N} n. 
(MPCE_{n}+\epsilon_n)}{\sum_n^{N} (MPCE_n+\epsilon_n) }\right) 
~~~~~~~~~~~\mbox{(from Eq.~\ref{Eqn:mpce=mpece}),} \nonumber \\
		&=& N.\left(\frac{\sum_n^{N} n. MPCE_{n}+ \sum_n^{N} n. \epsilon_n}{\sum_n^{N} MPCE_n+\sum_n^{N} \epsilon_n }\right).
\label{Eqn:mpece}
\end{eqnarray}		
Note that by similar logic, we have
\begin{equation}		
X^{MPCE}= N.\left(\frac{\sum_n^{N} n. MPCE_{n}}{\sum_n^{N} MPCE_n }\right).
\label{Eqn:mpce1}
\end{equation}
In Appendix~\ref{subsec:appendix_ineq} we provide a heuristic argument showing 
that given the ranking of expenditure in Eq.~\ref{Eqn:reorderE}, we have 
\begin{equation}
X^{MPECE}\ge X^{MPCE}. 		
\end{equation}
We need two more conditions viz. sufficient dispersion in the expenditure and less dispersion in the $\epsilon_n$ terms, both of which should hold in the data.	
Plugging the above inequality back in the equation for Gini coefficient 
(Eq.~\ref{Eqn:gini_expand}), we see that for any island $k$
\begin{equation}
G_k^{MPCE} \ge G_k^{MPECE},
\end{equation}
implying bigger inequality for state-wise comparisons. Since $k$-index is also highly correlated with Gini coefficient, the data shows similar features there as well.

\section{Dynamical features of inequality}
\label{sec:dynineq}
Ref.~\cite{Kuznets;55} presented a proposition that economies undergoing 
economic evolution shows an increasing trend in inequality initially before
a downward pressure builds on it which brings it down. Such an inverted `U'-shaped profile of inequality with respect to average income is known as the Kuznets' curve.
Although the actual time path followed by inequality as a function of per capita income is much more complex than the one originally proposed by Kuznets, it provides a basic intuitive understanding of
dynamics of inequality. However, this issue has been controversial as later 
research suggested that substantial inequality actually affects growth making 
the causal relationship less robust than it seems. In the same vein, 
Ref.~\cite{Deininger;97} shows that by itself economic growth may not directly 
contribute to the distributional outcomes adversely. Thus from a policy 
perspective this may not be a major factor. However,
one thing that is repeatedly seen is that some inequality seems natural 
companion of rapid growth. Redistributive policies are also seen to have 
ambiguous effects on growth. Inequality might grow even at a later stage of 
development. But evidence has been mixed~\cite{Atkinson;02}.
See also Ref.~\cite{angle2009kuznets} for a theoretical and 
empirical investigation of this mechanism.
\begin{figure}
 \includegraphics[width=\linewidth]{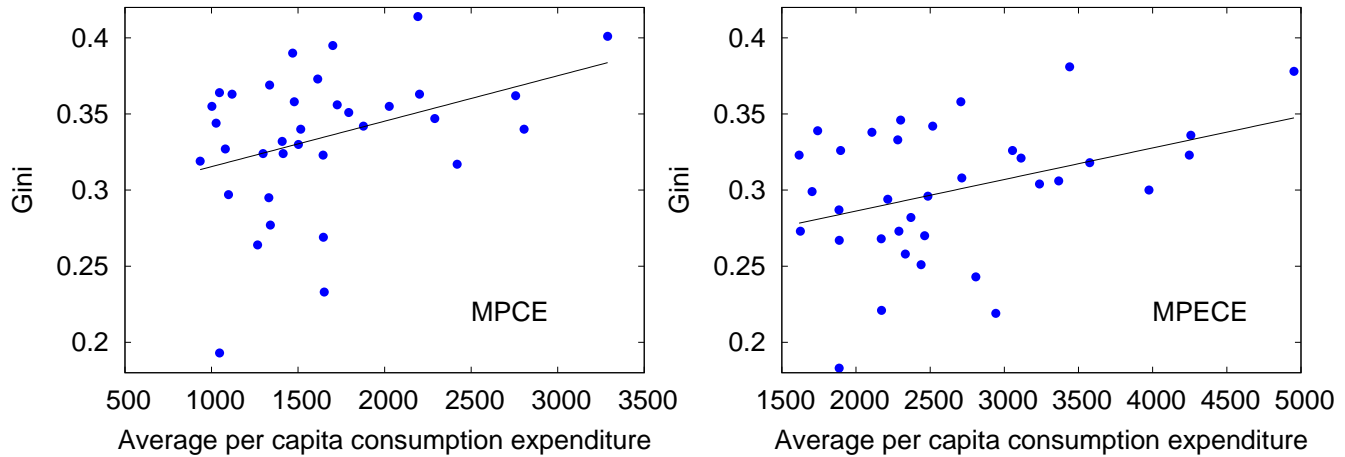}
\caption{Variation of Gini coefficient with average per capita consumption 
expenditure for MPCE and MPECE respectively. 
The regression curves are straight lines $Y=mX+c$, with $Y$ as Gini coefficient 
and $X$ as average per capita consumption expenditure.
For MPCE, $m=2.986 \times 10^{-5}$ and $c=0.285$ and 
for MPCE, $m=2.074 \times 10^{-5}$ and $c=0.245$.
An upward trend in absolute inequality measure (Gini) is clearly visible in 
both cases with increasing consumption. Note also that consumption is a decent 
proxy for income. Thus effectively this diagram is showing the first part of 
the celebrated Kuznets' curve.}
 \label{fig:growth}
 \end{figure}

What is of importance to us is the idea that in the initial stages of growth and development, the countries which are on an average more prosperous will
be somewhat more unequal than their less prosperous counterparts. Even though it was originally proposed as a time-series idea, we can easily adapt it to a multi-country set-up.
One can think of all countries following the same growth path over time (for 
example, in the spirit of the basic neo-classical growth model, 
also known as Solow model~\cite{Acemoglu-growth;09}). Thus a more prosperous 
country is nothing but what a poor country will become in future. Here we are 
making gross simplification in ignoring the roles of 
institutions~\cite{Acemoglu-nations-fail;13}. What we gain is
a framework to make comparisons between multiple economic entities at the same 
time. Noting that almost all of the Indian states are substantially large (for 
example, GDP of Hungary was  \$137.1 billion in 
2014~\cite{Worldbank-Hungary;14} and the same for the state Maharashtra is 
\$234.3 billion in the same year \cite{Wiki-Maharashtra;14}; data converted in 
to dollars based on the current exchange rate),
we can do a cross-state comparison. Fig.~\ref{fig:growth} shows state-wise 
consumption inequality versus average consumption plot. 
The regression fit is a liner growth $Y=mX+c$, where $Y$ is the Gini 
coefficient and $X$ is the average per capita consumption expenditure.
For MPCE, $m=2.986 \times 10^{-5}$ and $c=0.285$ and 
for MPECE, $m=2.074 \times 10^{-5}$ and $c=0.245$.

A very interesting feature of the plot is its apparent adherence to the basic proposition of Kuznets that in the initial growth periods more prosperous states should be more unequal.
In case of India, Ref.~\cite{kanbur2012does} presents arguments for and against 
the finding of such inequality dynamics. Even though cross-sectional estimates 
refute the claim, time-series estimates show some support to it. In a general 
context, Ref.~\cite{chakrabarti2010inequality}
show how such a feature may arise in a very simple dynamic model depending on relative strength of market correlations to average savings rate.
Further study on the evolution of inequality would be able to shade light on how 
far the data agree with the theory. As has been emphasized by 
Ref.~\cite{Atkinson;02}, inequality might increase even in developed economies 
depending on policies and broadly, institutional factors, it is a non-trivial 
task to try to predict which path will Indian economy take, both at the 
aggregate level and at the state level.

\section{Interpreting the results}
\label{sec:interpret}
In this section, we present a basic model to understand the empirical findings. The basic point we make here is that consumption profiles across the states can be invariant with respect
to average income or other factors. Another important point is that inequality in general increases as average consumption (which is highly correlated with income) increases.  

Suppose the economy consists of $J$ islands where each island can be identified with a country or in the present context, a state in India. There is a continuum of agents $I$ in every island $j \in J$. For simplicity, assume that $I=[0,1]$. Each agent is indexed by $i\in I$. Time is discrete. At every point of time, the agents are endowed with unit labor which are differentiated across workers. 
They do not value leisure and hence can provide labor inelastically.

There is a continuum of firm in every island which can combine labor to produce 
output with a standard production function. The aggregate level of production 
in island $j$ is
\begin{equation}
Y_j=z_j[\int_{I}\xi_{ij}^{\rho}d i]^{1/\rho}
\label{eqn:pdn_fn}
\end{equation}
where $z_j$ is the island-specific state of technology, $\xi_{ij}$ is the labor endowment of the $i$-th agent in the $j$-th state and $\rho$ is a parameter in the production function.
Labor endowment is a constant which is normalized to 1,
\begin{equation}
\xi_{ij}=1.
\label{Eqn:labor_endowment}
\end{equation} 
We will assume that the technology differs across islands and hence generally, $z_j\ne z_{j'}$ where $j,j'\in J$.
 
Firms maximize profit,
\begin{eqnarray}
\pi_j&=&P_jY_j-\int_I w_{ij}\xi_{ij}di\nonumber\\
     &=&z_jP_j[\int_{I}\xi_{ij}^{\rho}d i]^{1/\rho} -\int_I w_{ij}\xi_{ij}di
		\label{eqn:profit_fn}
\end{eqnarray}
where $P_j$ is the price of the final good and $w_j$ is the wage rate. In principle, we can normalize the general price level to unity without any loss of generalization.
The second equality can be obtained substituting the production function 
(Eq.~\ref{eqn:pdn_fn}) in the profit function (Eq.~\ref{eqn:profit_fn}) above.
By choosing optimal amount of labor, the firms arrive at the following first order condition:
\begin{equation}
\xi_{ij}=\left(\frac{P_jz_j}{w_{ij}}\right)^{\frac{1}{1-\rho}}\frac{Y_j}{z_j}.
\label{Eqn:labor_dd}
\end{equation}
It can be shown that the general price level is nothing but an aggregate function of wage rates,
\begin{equation}
P_j=z_j^{-1}[\int_I w_{ij}^{\frac{\rho}{1-\rho}} di]^{\frac{1-\rho}{\rho}}.
\end{equation}
Given the labor demand equation (Eq.~\ref{Eqn:labor_dd}), the equilibrium wage 
rate $w_{ij}$ can be found out by labor market clearing conditions (recall 
from Eq.~\ref{Eqn:labor_endowment} that total endowment is 1). 
Thus in equilibrium, GDP of the $j$-th island is given by
\begin{equation}
Y_j=z_j.
\end{equation}

Since the data is on household consumption expenditure, we have to discretize the model. Assume that any generic household $h$ is a set of agents $s_h=[p_h,p^h)$)and there are $N$ ($N\ge 1$) number of households in each island $j\in J$. We will denote the set of households in each island by $N$ as well. For the first household, $s_{1}=[0,p_1)$ and for the last household $N$, $s_{N}=[p_{N},1]$ in each island.
To maintain clarity, assume that any agent can belong to only one household. Thus,
\begin{equation}
\cup_{h\in N}s_h=I
\end{equation}
where we denote the set of households in a island by $N$.

In island $j\in J$, total production $Y_j$ is divided among $N$ households to consume. Clearly the contribution ($w_{hj}$) of the $h$th household is
\begin{equation}
w_{hj}=z_j\int_{s_h}\phi(\xi_{ij})di
\end{equation}
where $\phi(\xi_{ij})$ denotes the distribution of labor endowment on agents.
Finally to split the pie i.e. the total production $Y_j$, we assume that the households derive utility from consumption and have a bargaining power proportional to their
contribution. This is essentially a Nash bargaining situation. The resultant income distribution would be given by a vector which we denote by
$\{v_{hj}\}_{h\in N}$.

\subsection{Absolute inequality}
Given the income profile $\{v_{hj}\}_{h\in N}$ which is also the consumption profile as there no savings, we can easily pin down the level of absolute inequality. The easiest way to do it would be to consider
the standard deviation,
\begin{equation}
\sigma_{j}=\sqrt{\frac{1}{N}\sum_{h\in N}(v_{hj}-\langle v_{hj}\rangle)^2}
\label{Eqn:sdev_ineq}
\end{equation}
where $\langle .\rangle$ denotes expectation operator. Similarly, we could compute the Gini coefficient or the $k$-index.
Note that this explains the effects of having better technology on inequality which is explored below. Here, it should be mentioned that the way this model is set up, there is no
difference between income and consumption. Thus any income shock will be 
translated in to an aggregate consumption shock given the perfect sharing 
mechanism within each family. In reality that is typically not the 
case~\cite{Blundell;08,Perri;05}. However, we retain this assumption as it 
makes the model very simple and for present purpose, it suffices to assume no 
asset market and full within-family insurance.

\subsection{Comparison between state-level inequalities}
Consider two islands 1 and 2. Island 1 has a benchmark level of technology, $z_1=1$. Island 2 has a better technology, $z_2>z_1$.
Then it is easy to show that if all other things remain identical across the islands (productivity distributions, household distributions and production functions), then the income/consumption profile in island 2
is just a blown up version of the same in island 1. It is easy to show that in 
terms of standard deviation, inequality goes up (see Eq.~\ref{Eqn:sdev_ineq}). 
Thus in terms of
variance or standard deviation, the model explains increasing inequality. The basic reason is that these indexes are not scale invariant.

It is also easy to see that upon such normalization, these two consumption 
profiles coincide. This explains the invariance in consumption distribution upon 
normalization (Fig.~\ref{fig:states}).
Any measure of relative inequality that is scale invariant i.e. shows zero-degree homogeneity, should be unaffected by such a change.
Essentially this refers to the idea that relative inequality across islands should remain unchanged with respect to proportional change in income profile. This potentially presents a problem because
in this case the Lorenz curve would not change implying that the Gini coefficient will also not change.

Hence, we introduce one more ingredient. Along with better technology ($z_2>z_2$) which increases the size of the pie, let us assume that the households receive stochastic
endowments $e_{hi}\sim f(0,\sigma_i)$ in island $i\in \{1,2\}$ where $f(.)$ is probability density function with well defined moments. For sake of normalization, let us assume that $\sigma_1=0$ and $\sigma_2>\sigma_1$. Thus we get two income profiles. For island 1, the profile remains unchanged
$\{v'_{h1}\}_{h\in N}=\{v_{h1}\}_{h\in N}$ and $\{v'_{h2}\}_{h\in N_2}=\{v_{h2}+e_{h2}\}_{h\in N}$. 

Let us denote the means of these two series by $\langle v_1\rangle$ and $\langle v_2\rangle$. Then the two normalized series $\{v'_{h1}/\langle v_1 \rangle\}$ and $\{v_{h2}/\langle v_2 \rangle\}$  
have the same mean but the later is a mean-preserving spread of the former which 
immediately implies that $\{v'_{h1}/\langle v_1 \rangle\}$ stochastically 
dominates $\{v'_{h1}/\langle v_1 \rangle\}$  in the second 
order~\cite{Satya;09}. Therefore, the following inequality holds in terms of the 
cumulative density functions $F(.)$,
\begin{equation}
\int_{-\infty}^x[F_2(v'')-F_1(v'')]dv'' \ge 0 ~~\mbox{$\forall$ $x$}
\end{equation}
with strict inequality at some $x$ where $v''=v/\langle v\rangle$.
This immediately implies that $F_1 $ Lorenz dominates $F_2$ (see for example, 
Ref.~\cite{Satya;09}) and hence in terms of Gini coefficient $G$,
\begin{equation}
G_2>G_1.
\end{equation}
This explains the upward trend of inequality with respect to per capita consumption.

\section{Summary and conclusion}
\label{sec:summary}
In this paper, we have studied spatial invariance of inequality in case of 
India. We have analyzed data from 35 states and union territories.
The data is also available for different castes, religious adherences and 
urban-rural divide.
The main finding of this paper is that under suitable normalization the distributions collapse to a single distribution. This sheds light on the static features of inequality. In particular, it means that state-wise differences in inequality may arise from the differences in average incomes.
The spread seems to be fairly constant when that effect is taken away. 
The lognormal fit of the bulk and power law fit of the tail of the normalized 
distributions agree with the existing literature.

The generic form of the income distribution is given by a 
lognormal/gamma bulk and a power law tail. We show here that it is the same for 
consumption. However, the Pareto exponent is much larger for consumption 
compared to the income data, reflecting lower consumption inequality than 
income. By and large, this finding seems to be true in many empirical works. 
For example, Ref.~\cite{Perri;05} analyzes cross-sectional income 
and consumption data for U.S. and shows that income inequality increased 
substantially during the period 1980-2003 but consumption inequality did 
increase only by a small amount, reflecting the idea that consumption is less 
volatile than income at any given point of time. However, it is difficult to pin 
down an exact relationship between these two measures due to multiple external 
factors that might affect both consumption and income. A similar argument holds 
true for wealth as well.

Next we study the growth-inequality nexus and show that usually a higher level of prosperity is associated with a higher level of inequality. Given that India is on the first part of its growth track, this finding is in almost exact agreement with the basic statement of Kuznets' curve.
A brief model is presented to elucidate the idea that a growing pie may due to a better level of technology or riskier projects or more generally, a combination of them. This leads to higher inequality.

This is primarily showing the presence of universality in terms of inequality 
during the growth process of countries. Further theoretical work would help to 
explain the causal relationship between growth and rise in inequality, if any. 
It is also noteworthy that we have considered spatial features only (physical 
space as in states or in the parameter space of caste, religion or 
urban-rural divide). The temporal dimension has not been considered which 
should show both state-wise and aggregate evolution of inequality. However,
that lies beyond the scope of the present work. 
  
\appendix
\section{Data Tables}
\label{subsec:appendix}

Below we present compiled data on state wise inequality in 
Table~\ref{tab:per_cap_exp_gini_1}. Both the Gini coefficient and the $k$-index 
have been presented for all states along with average income ($\langle 
x\rangle$).

\begin{table}[h]
\begin{tabular}{|c|c|c|c|c|c|c|c|c|}
\hline

Index & State & \#Household & \multicolumn{3}{c|}{MPCE}   & 
\multicolumn{3}{c|}{MPECE}  \\ \cline{4-9} 
&  &  & $\langle x \rangle $ & Gini & $k$ & $\langle x \rangle $ & Gini & $k$  
\\ \hline 
1& Jammu \& Kashmir & 2726 & 1340.20 &  0.277 & 0.598 & 2332.55 &  0.258 & 
0.591 \\ \hline
2& Himachal Pradesh & 2043 & 1726.27 &  0.356 & 0.628 & 2713.30 &  0.308 & 
0.609 \\ \hline
3& Punjab & 3117 & 1877.41 &  0.342 & 0.623 & 3113.03 &  0.321 & 0.616 \\ 
\hline  
4& Chandigar & 305 & 3288.60 &  0.401 & 0.648 &  4952.62 &  0.378 & 0.639 \\ 
\hline
5& Uttaranchal & 1780 & 1413.57 &  0.324 & 0.615 & 2289.23 &  0.273 & 0.596 \\ 
\hline
6& Haryana & 2620 & 1792.95 &  0.351 & 0.625 & 3054.84 &  0.326 & 0.616 \\ 
\hline 
7& Delhi & 957 & 2805.90 &  0.340 & 0.622 & 4246.83 &  0.323 & 0.619 \\ \hline 
8& Rajasthan & 4138 & 1408.53 &  0.332 & 0.618 & 2370.31 &  0.282 & 0.599 \\ 
\hline
9& Uttar Pradesh & 8993 & 1080.05 &  0.327 & 0.616 & 1885.53 &  0.287 & 0.601 
\\ \hline
10& Bihar & 4568 & 934.70 &  0.319 & 0.614  & 1624.77 &  0.273 & 0.596 \\ \hline
11& Sikkim & 768 & 1644.50 &  0.323 & 0.620 & 2438.57 &  0.251 & 0.591 \\ 
\hline 
12& Arunachal Pradesh & 1642 & 1297.71 &  0.324 & 0.616 & 2213.81 &  0.294 & 
0.604 \\ \hline
13& Nagaland & 1024 & 1651.50 &  0.233 & 0.583  & 2942.54 &  0.219 & 0.579 \\ 
\hline
14& Manipur & 2558 & 1046.59 &  0.193 & 0.566 & 1886.58 &  0.183 & 0.564 \\ 
\hline 
15& Mizoram & 1528 & 1646.45 &  0.269 & 0.597  & 2808.10 &  0.243 & 0.588 \\ 
\hline
16& Tripura & 1856 & 1331.25 &  0.295 & 0.607 & 2170.16 &  0.268 & 0.596 \\ 
\hline
17& Meghalaya & 1272 & 1266.10 &  0.264 & 0.594  & 2171.71 &  0.221 & 0.579 \\ 
\hline
18& Assam & 3448 & 1098.20 &  0.297 & 0.607 & 1887.38 &  0.267 & 0.597 \\ \hline
19& West Bengal & 6324 & 1335.42 &  0.369 & 0.635 & 2107.57 &  0.338 & 0.622 \\ 
\hline
20& Jharkhand & 2751 & 1026.32 &  0.344 & 0.624 & 1704.10 &  0.299 & 0.607 \\ 
\hline
21& Orissa & 4031 & 1001.90 &  0.355 & 0.627 & 1616.64 &  0.323 & 0.615 \\ 
\hline
22& Chattisgarh & 2230 & 1045.88 &  0.364 & 0.631 & 1741.39 &  0.339 & 0.622 
\\ \hline
23& Madhya Pradesh & 4705 & 1118.79 &  0.363 & 0.630 & 1896.47 &  0.326 & 0.616 
\\ \hline
24& Gujarat & 3425 & 1501.83 &  0.330 & 0.620 & 2484.72 &  0.296 & 0.607 \\ 
\hline
25& Daman \& Diu & 128 & 2026.93 &  0.355 & 0.629 & 3236.79 &  0.304 & 0.610 \\ 
\hline
26& D \& N Haveli & 192 & 1515.47 &  0.340 & 0.626 & 2462.95 &  0.270 & 0.599 
\\ \hline
27& Maharashtra & 8005 & 1700.64 &  0.395 & 0.643  & 2706.95 &  0.358 & 0.628 
\\ \hline
28& Andhra Pradesh & 6889 & 1613.82 &  0.373 & 0.635 & 2517.41 &  0.342 & 0.623 
\\ \hline
29& Karnataka & 4074 & 1468.72 &  0.390 & 0.641 & 2300.90 &  0.346 & 0.624 \\ 
\hline
30& Goa & 447 & 2419.38 &  0.317 & 0.611 & 3975.30 &  0.300 & 0.605 \\ \hline
31& Lakshadweep & 184 & 2201.62 &  0.363 & 0.633 & 3366.33 &  0.306 & 0.611 \\ 
\hline
32& Kerala & 4455 & 2192.06 &  0.414 & 0.648 & 3440.41 &  0.381 & 0.635 \\ 
\hline
33& Tamil Nadu & 6639 & 1478.30 &  0.358 & 0.630 & 2281.49 &  0.333 & 0.621 \\ 
\hline
34& Pondicherry & 576 & 2289.77 &  0.347 & 0.625 & 3575.46 &  0.318 & 0.615 \\ 
\hline 
35& A \& N Island & 559 & 2757.06 &  0.362 & 0.632 & 4257.64 &  0.336 & 0.622 
\\ \hline
\end{tabular}
\caption{Average per capita consumption expenditure $\langle x \rangle$, Gini 
and $k$-indices. Data is shown for MPCE \& MPECE.}
\label{tab:per_cap_exp_gini_1}

\end{table}

In the following we also present, caste, religion and location-based inequality measures along with their average prosperity.

\begin{table}[t]
\begin{tabular}{|c|c|c|c|c|c|c|c|c|}
\hline

Index & Caste & \#Household & \multicolumn{3}{c|}{MPCE}   & 
\multicolumn{3}{c|}{MPECE}  \\ \cline{4-9} 
&  & & $\langle x \rangle $ & Gini & $k$ & $\langle x \rangle $ & Gini & $k$  
\\ \hline 
1& ST & 12928 & 1199.02 &  0.339 & 0.621   & 2031.68 &  0.313 & 0.611  \\ \hline
2& SC & 16181 & 1114.59 &  0.329 & 0.617  & 1853.54 &  0.294 & 0.603  \\ \hline
3& OBC & 37872 & 1316.96 &  0.352 & 0.626   & 2164.61 &  0.315 & 0.612  \\ 
\hline
9& General & 33912 & 1817.78 &  0.383 & 0.639  & 2902.85 &  0.346 & 0.625  \\ 
\hline
\end{tabular} 
\caption{Average per capita consumption expenditure $\langle x \rangle$, Gini 
and $k$-indices for different castes. Data is shown for MPCE \& MPECE.}
\end{table}
\begin{table}[t]
\begin{tabular}{|c|c|c|c|c|c|c|c|c|}
\hline

Index & Religion & \#Household & \multicolumn{3}{c|}{MPCE}   & 
\multicolumn{3}{c|}{MPECE}  \\ \cline{4-9} 
& & & $\langle x \rangle $ & Gini & $k$ & $\langle x \rangle $ & Gini & $k$  \\ 
\hline 
1& Hindu & 76949 & 1429.07 &  0.376 & 0.636   & 2311.01 &  0.339 & 0.621  \\ 
\hline  
2& Muslim & 12439 & 1245.50 &  0.350 & 0.624 & 2117.70 &  0.310 & 0.610  \\ 
\hline 
3& Christian & 6948 & 1688.70 &  0.355 & 0.628   & 2799.80 &  0.320 & 0.614  \\ 
\hline 
4& Other religion & 4598 & 1718.93 &  0.366 & 0.632   & 2857.81 &  0.337 & 
0.621  \\ \hline  
\end{tabular}
\caption{Average per capita consumption expenditure $\langle x \rangle$, Gini 
and $k$-indices for different religions. Data is shown for MPCE \& MPECE.}

\end{table}
\begin{table}[t]
\begin{tabular}{|c|c|c|c|c|c|c|c|c|}
\hline

Index &  & \#Household & \multicolumn{3}{c|}{MPCE}   & 
\multicolumn{3}{c|}{MPECE}  \\ \cline{4-9}
& & & $\langle x \rangle $ & Gini & $k$ & $\langle x \rangle $ & Gini & $k$  \\ 
\hline 
0& Rural & 41828 & 1865.76 &  0.384 & 0.638   & 2942.37 &  0.347 & 0.625  
\\ \hline  
1& Urban & 59129 & 1134.62 &  0.313 & 0.610  & 1923.88 &  0.286 & 0.600  \\ 
\hline 
\end{tabular} 
\caption{Average per capita consumption expenditure $\langle x \rangle$, Gini 
and $k$-indices for different geographic locations. Data is shown for MPCE \& 
MPECE.}

\end{table}
\section{Relative inequality}
\label{subsec:appendix_ineq}
Recall that (Eq.~\ref{Eqn:mpece} and \ref{Eqn:mpce1}) we have
\begin{equation}
X^{MPECE} = N.\left(\frac{\sum_n^{N} n. MPCE_{n}+ \sum_n^{N} n. \epsilon_n}{\sum_n^{N} MPCE_n+\sum_n^{N} \epsilon_n }\right),
\end{equation}		
and
\begin{equation}		
X^{MPCE}= N.\left(\frac{\sum_n^{N} n. MPCE_{n}}{\sum_n^{N} MPCE_n }\right).
\end{equation}
For simplicity, let us assume that $\epsilon_n\approx \epsilon~\forall n$. Then
\begin{equation}
X^{MPECE} = N.\left(\frac{\sum_n^{N} n. MPCE_{n}+  \left(\frac{N(N+1)}{2}\right).\epsilon}{\sum_n^{N} MPCE_n+N. \epsilon }\right).
\end{equation}		
Let us denote the relationship between $X^{MPECE}$ and $X^{MPCE}$ by $\mathcal{R}$. We want to check if $\mathcal{R}$ is $\ge$ or $\le$.
We can write,
\begin{equation}
\left(\frac{\sum_n^{N} n. MPCE_{n}+  \left(\frac{N(N+1)}{2}\right).\epsilon}{\sum_n^{N} MPCE_n+N. \epsilon }\right)~ \mathcal{R} ~ \left(\frac{\sum_n^{N} n. MPCE_{n}}{\sum_n^{N} MPCE_n }\right)
\end{equation}		
which can be simplified to
\begin{equation}
\frac{N(N+1)}{2}.\epsilon \left( \sum_n^{N} MPCE_n\right) ~ \mathcal{R} ~ N\epsilon \left( \sum_n^{N} n.MPCE_n \right)
\end{equation}
or
\begin{equation}
\frac{(N+1)}{2}.\left( \sum_n^{N} MPCE_n \right) ~ \mathcal{R} ~ \left( \sum_n^{N} n.MPCE_n  \right).
\end{equation}
The above expression can be rewritten as
\begin{equation}
\sum_n^{N} \left( \left( \frac{(N+1)}{2}-n\right).MPCE_{n}\right) ~\mathcal{R} ~ 0.
\end{equation}
From Eq.~\ref{Eqn:reorderE} (without loss of generalization, assuming that the 
family sizes are identical), the above relationship clearly shows that 
$\mathcal{R}$ is $\ge$ i.e.
\begin{equation}
\sum_n^{N} \left( \left( \frac{(N+1)}{2}-n\right).MPCE_{n}\right) \ge 0.
\end{equation}
This implies,
\begin{equation}
X^{MPECE}\ge X^{MPCE}.
\end{equation}


\bibliographystyle{unsrt}
\bibliography{expe}

\end{document}